# Enhanced production of coherent pulsed radiation at 125 nm: the route towards a tabletop VUV laser.

S. Chénais, S. Forget and M.-C. Castex

*Laboratoire de Physique des Lasers (LPL, CNRS), Institut Galilée, Université Paris 13, 93430 Villetaneuse, France.*

*e-mail : chenais@galilee.univ-paris13.fr*

**Abstract:** A novel approach is used to enhance by nearly two orders of magnitude the conversion efficiency of a 125 nm-coherent source, based on four-wave mixing in room-temperature mercury vapor. Saturation issues are observed and discussed.

## 1. Introduction

The need for compact, reliable and robust sources in the VUV region (100<λ<200 nm) is increasingly strong, due to its numerous emerging applications: we may mention nanopatterning and nanolithography, photochemical ablation of polymers [1], biochemistry, and more recently (exo-)biology and radioprotection. All these applications benefit from an efficient (and athermal) one-photon interaction of light with matter. In most cases, synchrotron radiation is used, and in some cases excimer lasers, but the latter suffer from weak coherence, fixed wavelength and troublesome maintenance. Nonlinear conversion of visible and IR lasers in vapors (crystals are not transparent enough in this domain) has been extensively studied since the 70's; in this context mercury vapor has emerged as one of the most efficient nonlinear media for the production of 120-185 nm radiation (recently a Hg-based source has been used to produce Lyman-α CW radiation for cooling antihydrogen atoms [2]). However, these sources involve up to three different spectrally narrow lasers, and mercury is heated at ~200 °C to make the atomic density large enough to ensure reasonable efficiency. Since deposition of opaque Hg on windows is an issue, very complex cell geometries have been proposed. The presence of hot mercury severely restricts the practical or industrial use of these sources, for both technical and ecological reasons. We presented in a previous paper a double resonance strategy to produce coherent 125-nm light with a single dye laser at 625.7 nm using a *room-temperature* Hg cell [3]. We present in this paper a way to enhance the conversion efficiency without heating the mercury, by locally increasing the atomic density (both spatially and temporally) thanks to an external laser focussed on the mercury pool (?) to provoke controlled vaporization.

## 2. Experimental setup

The VUV radiation at 125.14 nm is obtained by a four-wave mixing process shown schematically in fig. 1. The frequency $\omega_1$ of the dye laser (DCM+Rhodamine 640) is adjusted in order that its frequency-doubled output at $\omega_2 = 2\omega_1$ is resonant with the $6^1S_0 \rightarrow 7^1S_0$ two-photon transition of Hg. The colinear beams at $\omega_1$ and $\omega_2$ (2.6 mJ and 0.7 mJ incident onto the Hg cell, respectively) are tightly focussed to a ~40μm spot (in diameter) with a 280-mm focal length achromatic doublet. To demonstrate the principle of an enhancement, we chose as a vaporization source an ArF excimer laser (Neweks PSX-100, 3 mJ @193 nm, 4 ns pulse duration) because of the strong absorption coefficient of liquid Hg at this wavelength at normal incidence (70 %). The ArF beam was focussed onto the Hg surface with a 50-mm quartz lens. The VUV signal is measured by a calibrated CsI solar-blind photomultiplier. The intensities of red and UV beam were adjustable separately (see fig. 1.)

## 3. Results and discussion

As shown in fig.2a, the largest enhancement factor is obtained when the vaporization laser hits the surface ~1.5 μs before the UV and red beams. When the lossy elements (waveplates and polarizers) were removed, we obtained a maximum VUV energy of 2.5 μJ (300 W peak power inside the cell) corresponding to ~1MW/cm$^2$ and an enhancement factor of 6. But interestingly, this enhancement increases dramatically when both the UV and the visible intensities are decreased, as shown in figure 2b for the UV beam and attains here values >60 for very low UV fluences (for lower UV energies the VUV output without vaporization was too small to be measurable).

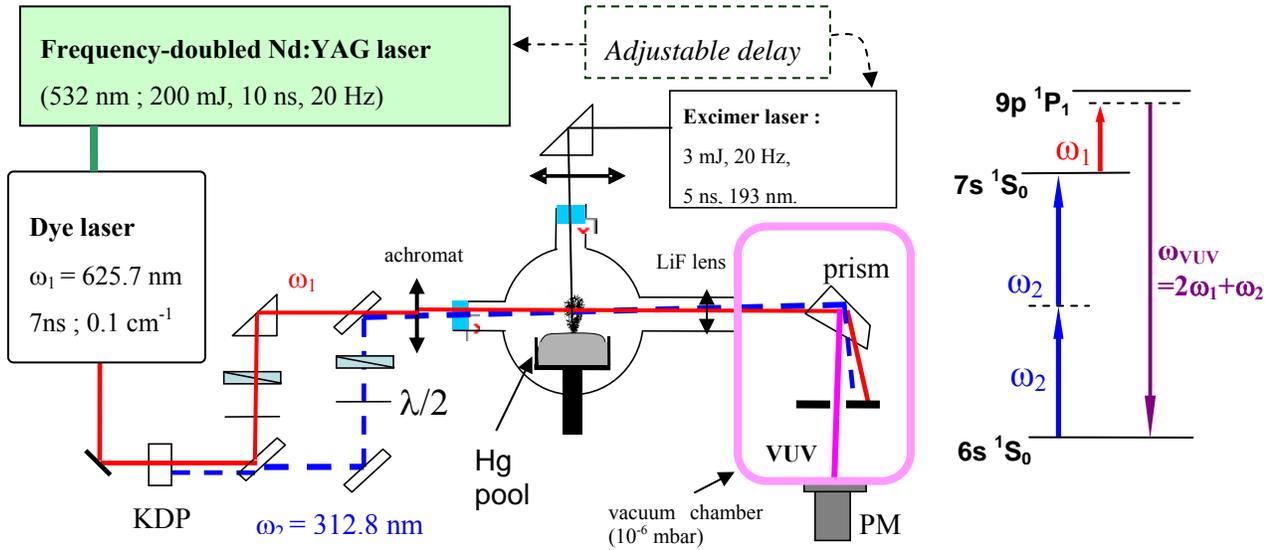

**Figure 1**: experimental setup (left) ; energy diagram (right)

This strong saturation proves that the vaporization does not only increase the Hg atomic density : the VUV yield is affected by a modification of phase matching conditions due to a significant saturation of the two-photon transition, as well as by the competition with other nonlinear effects (parametric oscillation, ASE…), as can be shown by a XUV spectrum analysis. This means that the tight focussing regime is not optimal any more when the vaporizing laser in present.

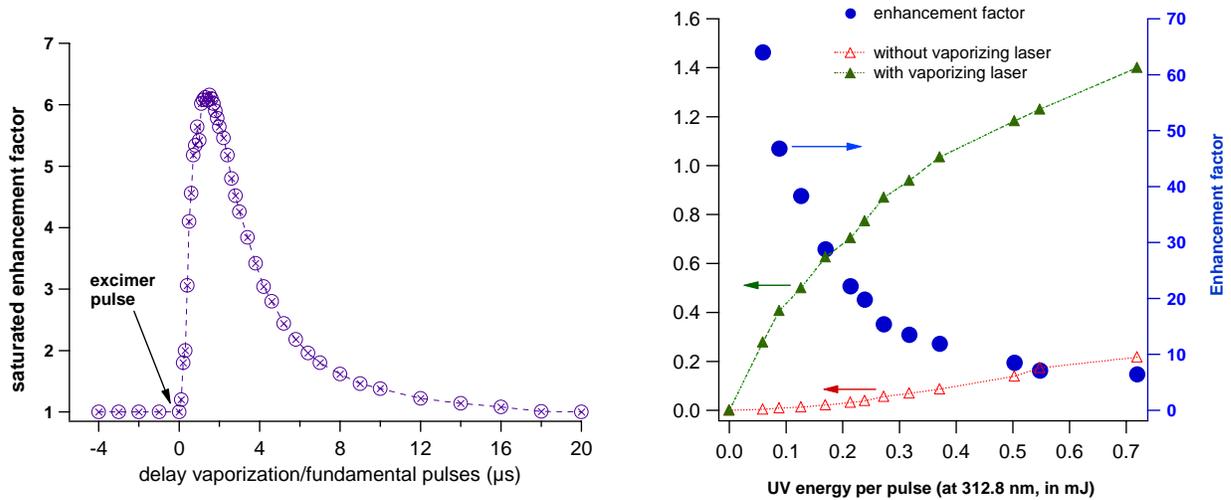

**Fig. 2.a** (left): saturated enhancement factor (for max. intensities at $\omega_1$ and $\omega_2$) vs. delay between vaporizing laser and fundamental beams; **fig. 2b** (right): VUV yield (left) and enhancement factor (right) vs. UV energy at constant energy at $\omega_1$ (2,6 mJ). The ArF energy is 3 mJ and the temporal shift (see fig. 2a) is fixed to 1.5 µs.

These first results show that an efficient source based on a room-temperature Hg cell is feasible using lower fluences *i.e.* larger beams. Provided that the lasers are replaced by state-of-the-art solid-state sources, these results are highly encouraging for the realization of a tabletop (surface < 2 m$^2$) VUV laser.